# Impact of microlens shape on the performance of Laser Guide Star wavefront sensors for ELT-class telescopes

Paul Rouquette*[b], Benoit Neichel[a] , Cédric Taïsir Héritier[c,a], Pierre Jouve[c], Anne Costille[a], Kjetil Dohlen[a], Thierry Fusco[c,a]
[a]Aix Marseille University, CNRS, CNES, LAM, Marseille France;
[b]Aix Marseille Univ, CNRS, Centrale Med, Institut Fresnel, Marseille, France
[c]DOTA, ONERA, 13661 Salon cedex AIR, France;
[c]Space ODT – Optical Deblurring Technologies[1]

## ABSTRACT

Laser Guide Star (LGS) adaptive optics systems on extremely large telescopes (ELTs) rely on Shack-Hartmann wavefront sensors (SHWFS) equipped with large-format microlens arrays. Manufacturing imperfections in the microlens surface profile degrade spot quality and reduce centroiding accuracy, yet this effect is rarely quantified in the context of full AO system performance. This paper presents a comprehensive characterization of the impact of microlens shape on LGS wavefront sensing, using the MORFEO/HARMONI LGS wavefront sensor design as the primary test case — results are directly applicable to any ELT-class or future large-telescope LGS instrument, including systems on the GMT and TMT. Starting from interferometric surface profile measurements of prototype microlenses, we derive the induced phase errors and compute the degradation of center-of-gravity (CoG) spot detection accuracy as a function of LGS elongation and flux. A real microlens degrades CoG variance by a factor of 1.8 to 2.4 compared to an ideal lens, equivalent to requiring approximately twice the photon flux to maintain the same measurement accuracy. This effect is shown to decrease with increasing LGS elongation, and to improve with higher microlens sag. The per-subaperture flux-loss model is then propagated into tomographic AO end-to-end simulations, where measurement redundancy across multiple LGS and MMSE reconstructor weighting partially mitigate the penalty, reducing the effective Strehl ratio flux loss to factors of 1.25–1.4. Experimental validation is provided with the LGS wavefront sensor optical bench prototype at LAM, which also enables full-array characterization in a single measurement. The methodology is broadly applicable to any Shack-Hartmann system operating in a low-flux regime.

## 1. INTRODUCTION

Laser Guide Star (LGS) adaptive optics (AO) systems are a cornerstone of future extremely large telescope instrumentation. By generating artificial sodium laser beacons at ~90 km altitude, these systems enable diffraction-limited performance over a large fraction of the sky. On the Extremely Large Telescope (ELT), multiple instruments will exploit LGS-based AO: MORFEO provides multi-conjugate AO (MCAO) correction for MICADO and HARMONI, while MOSAIC will develop 4 LGSWFS. Future instruments at the ELT, GMT, and TMT will similarly depend on LGS wavefront sensing. In all these systems, the LGS wavefront sensor is a Shack-Hartmann (SH) device composed of a large-format microlens array, in the case of MORFEO, a 68×68 double-sided array.

End-to-end AO simulations used to drive instrument design typically assume ideal microlenses. In practice, however, manufacturing tolerances cause the microlens surface profile to deviate from the ideal sphere, degrading the focal-plane PSF and reducing wavefront sensing accuracy. This effect is particularly critical in the low-flux regime characteristic of LGS operation on large telescopes, where photon budgets are tight and wavefront sensor noise directly limits AO performance. Several prototype microlens arrays have been manufactured for the MORFEO LGS wavefront sensor, and the main objective of this paper is to characterize the impact of their shape on AO performance and derive specifications for future production arrays. The results and methodology are applicable to any SH-based LGS system on an ELT-class telescope. A direct high-resolution simulation of all 4624 lenslets in a full tomographic AO simulation is computationally intractable. We therefore propose an efficient two-step approach: first, the effect of a single real microlens on spot centroiding is characterized analytically and numerically as a function of LGS elongation and flux; second, this per-subaperture result is propagated into a full AO end-to-end simulation by modeling microlens imperfections as an effective subaperture flux reduction. Experimental validation using the LGS wavefront sensor bench prototype at LAM completes the study.

*corresponding author: paul.rouquette@fresnel.fr

## 2. MICROLENS IMPACT ON CENTER-OF-GRAVITY MEASUREMENTS

The positions of SH spots in the focal plane are estimated by a center-of-gravity (CoG) algorithm, and the CoG variance serves as the primary figure of merit for detection accuracy.[34] In this section we compare CoG variance for ideal vs. real prototype microlenses as a function of LGS flux, elongation, and lens shape.

### 2.1 Optical model of a microlens

Surface profile measurements of prototype microlenses were acquired with a Wyko NT 9100 Optical Profiling System. A sphere is fitted to the measured surface; the optical path difference (OPD) introduced by the real lens relative to the ideal fitted sphere is:

$$\Delta = (n-1)(z_{real} - z_{ideal}) \tag{1}$$

where n is the refractive index of the lens material, and $z_{real}$ and $z_{ideal}$ are the real and ideal surface heights respectively. From this OPD the phase aberration φ is derived, and the PSF of each microlens is computed as the squared modulus of the Fourier transform of the complex pupil field. Figure 1 shows the measured altitude profile, the fitted sphere, and the resulting phase map. Figure 2 compares the PSFs of the ideal and real lenses: the real lens redistributes flux from the central peak toward the wings, degrading CoG precision.

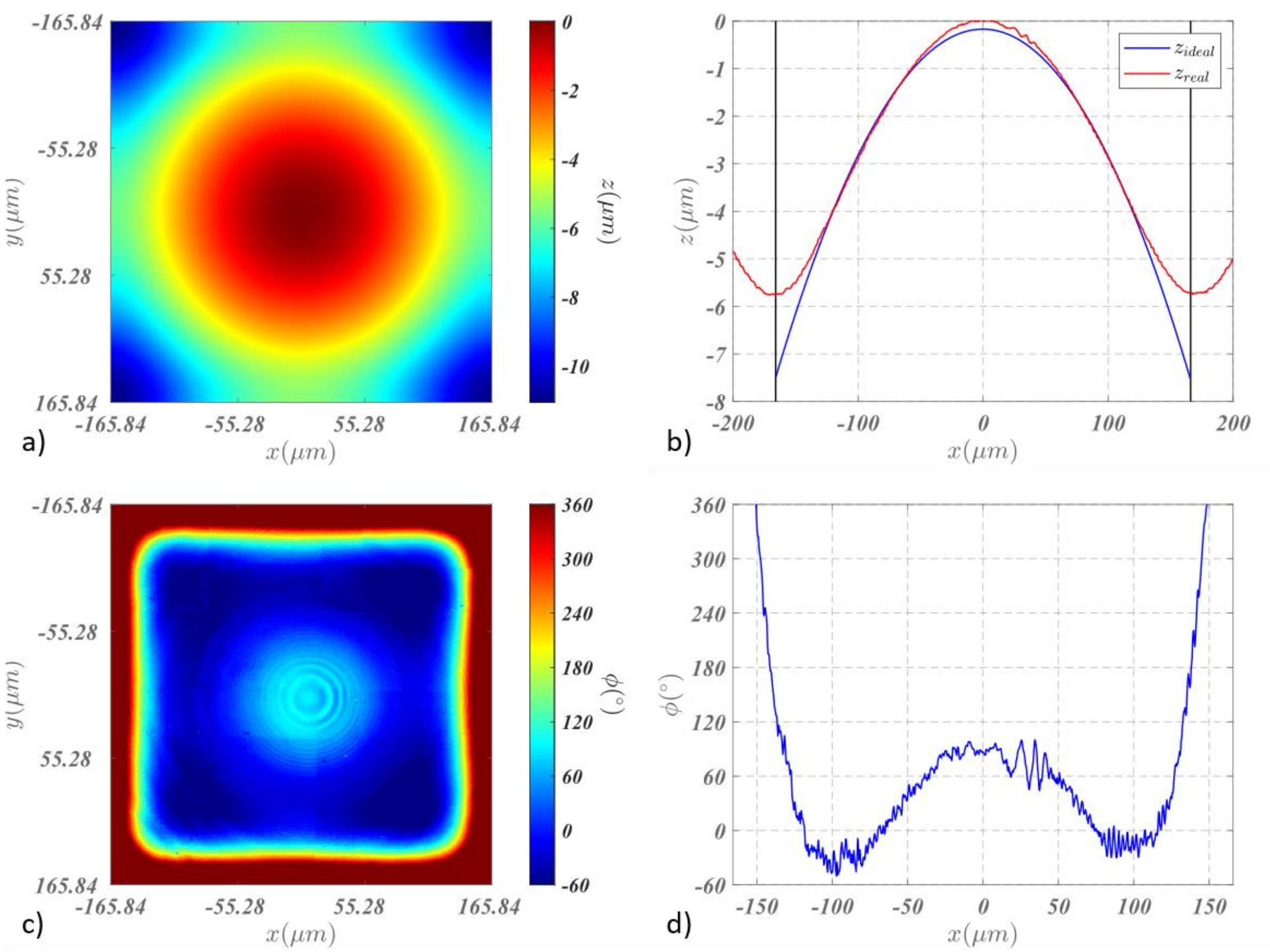

Fig 1. (a) Altitude profile of a prototype microlens; lens center at (x,y) = (0,0). (b) Altitude profile along x-axis (y = 0): measured surface (red) and fitted sphere (blue). (c) Phase difference introduced by the lens (radians). (d) Phase along the x-axis (y = 0).

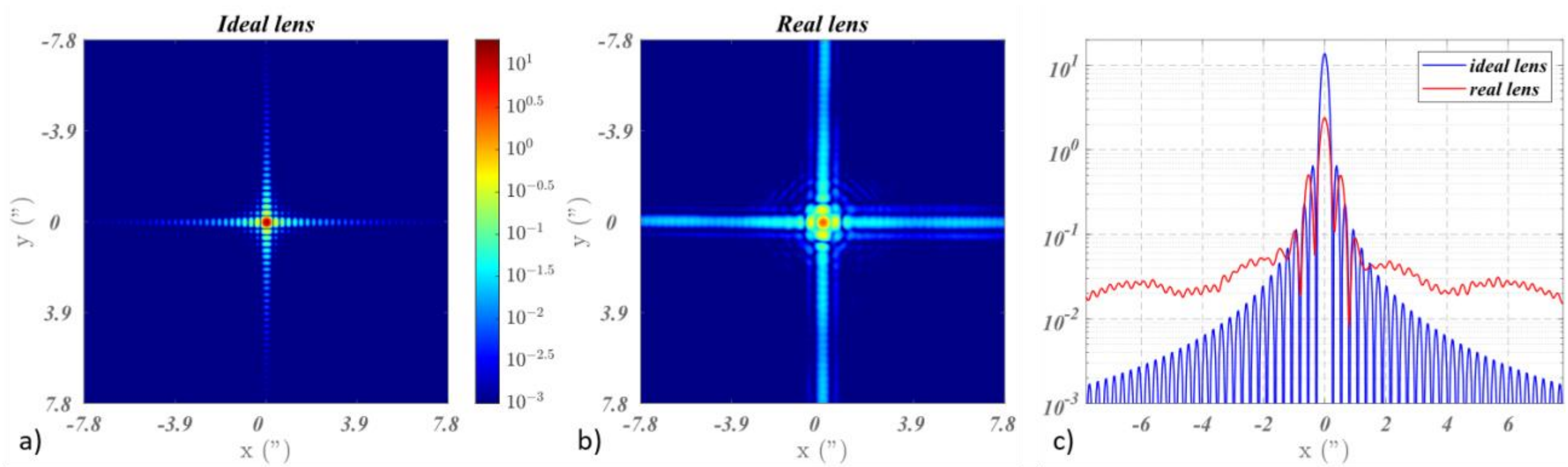

Fig 2. (a) PSF of the ideal lens. (b) PSF of the real lens. (c) Comparison of PSF profiles along the x-axis (y = 0). The real lens shifts flux from the central peak to the wings.

### 2.2 Light distribution on the detector

The light distribution on detector pixels is obtained by convolving the LGS intensity profile with the microlens PSF, then sampling with the detector pixel grid. The LGS is modeled as a uniform sodium layer profile (Fig. 3), generating spots of varying elongation depending on subaperture position relative to the launch telescope. Each subaperture is sampled by 14×14 pixels of 1.2 arcsec width.

In the non-elongated case (Fig. 4a–c), the real microlens produces a flux drop at the spot center and pronounced wings at the edges. In the elongated case (Fig. 4d–f), the impact is more diffuse, with a relatively uniform flux reduction across the subaperture. After pixel sampling (Fig. 5), these effects translate directly into degraded centroiding performance.

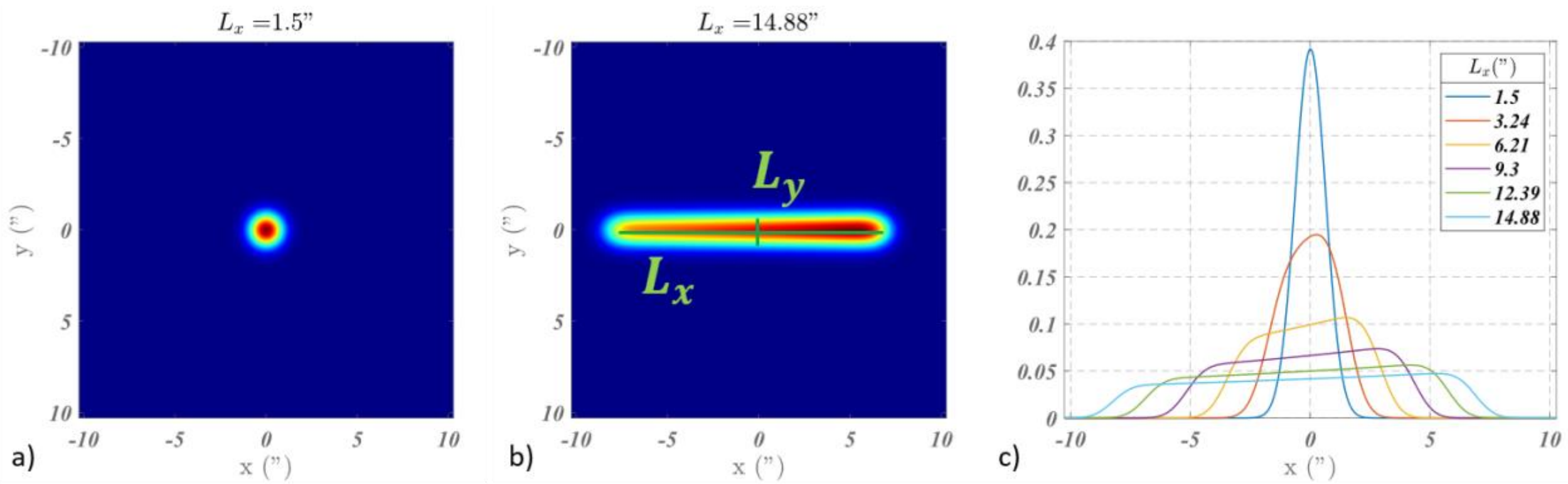

Fig 3. LGS intensity profiles used in simulations: (a) non-elongated spot ($L_x$ = 1.5″); (b) most elongated spot ($L_x$ = 14.88″); (c) profiles for several elongation values along the x-axis.

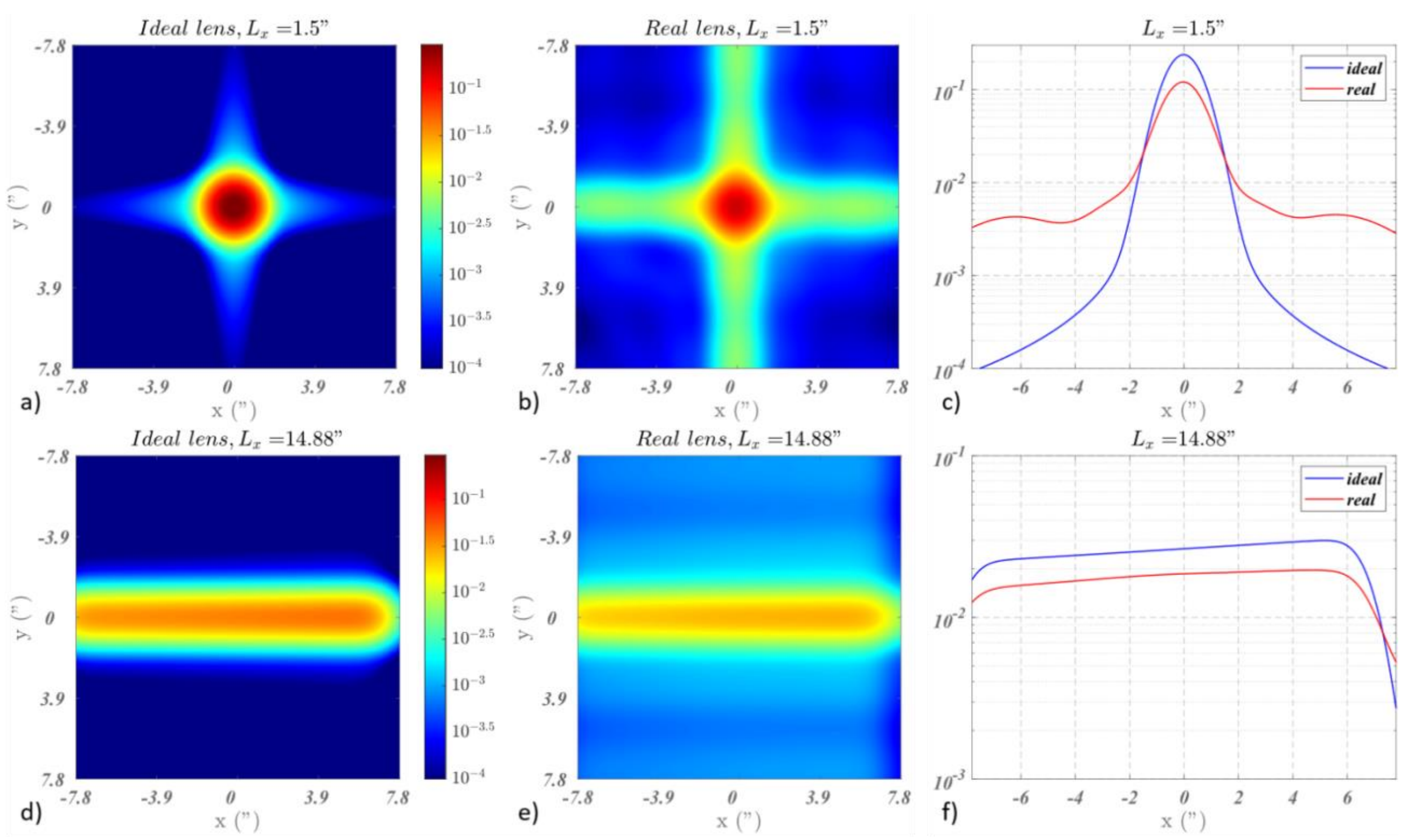

Fig 4. Light distribution on the detector before pixel sampling, for ideal and real lenses, with non-elongated (top row) and maximally elongated (bottom row) LGS.

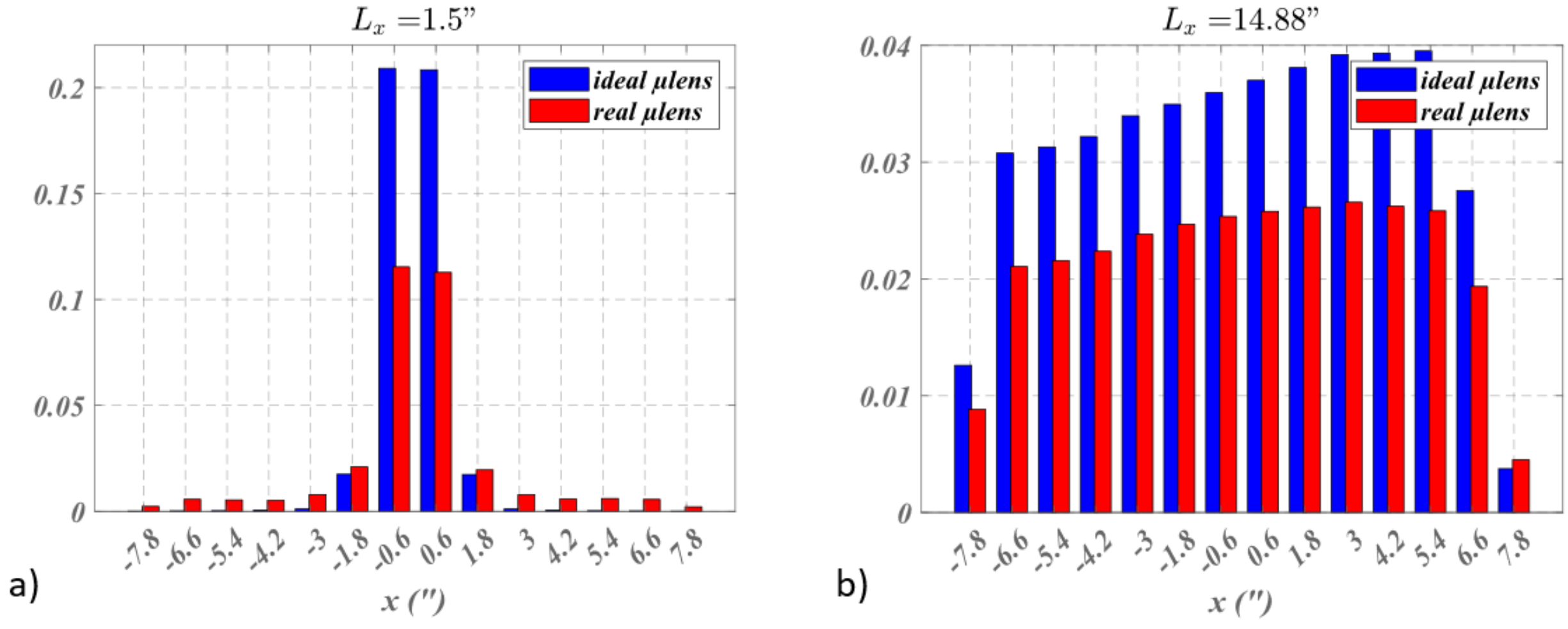

Fig 5. Sampled light distribution on detector pixels (14×14 per subaperture) for ideal and real lenses, non-elongated (a) and maximally elongated (b) LGS.

### 2.3 Centroiding study with weighted center of gravity

In this section we analyse the detection accuracy of the spot position on the detector in the presence of the two main sources of noise: photon noise (PN) and read-out noise (RON). By considering Gaussian spots, analytical expressions for the WCoG variance can be derived (Section 2.3.1), which are then used to confirm the numerical results in Sections 2.3.2–2.3.4.

### 2.3.1 Theoretical analysis

To detect the position of the spot, a weighted center of gravity (WCoG)[4] is used. It consists in multiplying the spot intensity by a Gaussian map and then performing a classical center-of-gravity measurement, leading to the following expression:

$$C_x = \frac{1}{b_x}\frac{\sum_{i,j} x_{ij} I_{ij} W_{ij}}{\sum_{i,j} I_{ij} W_{ij}} \text{ with } b_x = \frac{W_x^2}{W_x^2 + L_x^2}$$
$$C_y = \frac{1}{b_y}\frac{\sum_{i,j} y_{ij} I_{ij} W_{ij}}{\sum_{i,j} I_{ij} W_{ij}} \text{ with } b_y = \frac{W_y^2}{W_y^2 + L_y^2} \quad (2)$$

where $x_{ij}$, $y_{ij}$ are the pixel positions, $I_{ij}$, $W_{ij}$ are the associated spot intensity and weighting function. The terms $b_x$ and $b_\gamma$ are measurement biases introduced by the fact that the weighting function is centred on the detector center, not on the spot position (which is unknown a priori). These biases depend on the FWHM of the spot ($L_x$, $L_\gamma$) and the weighting Gaussian ($W_x$, $W_\gamma$) on both axes.

To compute the WCoG variance analytically, two hypotheses are made: (i) the main noise sources are photon noise (Poisson with mean N) and RON (centered Gaussian with standard deviation $\sigma_{RON}$);[34] (ii) both the spot I and the weighting function W are Gaussian:

$$I(x,y) = N\left(\sqrt{\frac{4\ln(2)}{\pi L_x^2}} e^{-\frac{4\ln(2)x^2}{L_x^2}}\right)\left(\sqrt{\frac{4\ln(2)}{\pi L_y^2}} e^{-\frac{4\ln(2)y^2}{L_y^2}}\right) \quad (3)$$

$$W(x,y) = \left(\sqrt{\frac{4\ln(2)}{\pi W_x^2}} e^{-\frac{4\ln(2)x^2}{W_x^2}}\right)\left(\sqrt{\frac{4\ln(2)}{\pi W_y^2}} e^{-\frac{4\ln(2)y^2}{W_y^2}}\right) \quad (4)$$

where N is the number of photons per subaperture per frame, and ($L_x$, $L_\gamma$), ($W_x$, $W_\gamma$) are the FWHM of the spot and weighting function on both axes.

Under these hypotheses, the WCoG variance is computed by separating the photon noise and RON contributions:[3]

$$\sigma_x^2 = \sigma_{x,PN}^2 + \sigma_{x,RON}^2 = \frac{1}{b_x^2}\frac{\iint x^2 W^2(x,y) I(x,y)dxdy}{(\iint W(x,y)I(x,y)dxdy)^2} + \frac{1}{b_x^2}\frac{\sigma_{RON}^2}{d^2}\frac{\iint x^2 W^2(x,y)dxdy}{(\iint W(x,y)I(x,y)dxdy)^2}$$
$$\sigma_y^2 = \sigma_{y,PN}^2 + \sigma_{y,RON}^2 = \frac{1}{b_y^2}\frac{\iint y^2 W^2(x,y) I(x,y)dxdy}{(\iint W(x,y)I(x,y)dxdy)^2} + \frac{1}{b_y^2}\frac{\sigma_{RON}^2}{d^2}\frac{\iint y^2 W^2(x,y)dxdy}{(\iint W(x,y)I(x,y)dxdy)^2} \quad (5)$$

Fixing the ratio of the weighting function width to the spot width (β = $W_x/L_x$ = $W_\gamma/L_\gamma$, typically β = 2), the variance expressions simplify to:

$$\sigma_x^2 = \frac{(\beta^2+1)^2}{\beta^4}\frac{L_x^2}{8\ln(2)N}\frac{(\beta^2+1)^2}{(\beta^2+2)^2} + \frac{(\beta^2+1)^2}{\beta^4}\frac{\sigma_{RON}^2}{d^2}\frac{\pi L_x^3 L_y}{128\ln^2(2)N^2}(\beta^2+1)^2$$
$$\sigma_y^2 = \frac{(\beta^2+1)^2}{\beta^4}\frac{L_y^2}{8\ln(2)N}\frac{(\beta^2+1)^2}{(\beta^2+2)^2} + \frac{(\beta^2+1)^2}{\beta^4}\frac{\sigma_{RON}^2}{d^2}\frac{\pi L_y^3 L_x}{128\ln^2(2)N^2}(\beta^2+1)^2 \quad (6)$$

with β = $W_x/L_x$ = $W_\gamma/L_\gamma$. Equation (6) generalises the formula of Thomas et al. (2006)[5] to handle asymmetric (elongated) spots: elongation on one axis also degrades centroiding on the perpendicular axis through the RON term. For large LGS elongations approaching the detector boundary, a finite-detector correction must be applied:

$$\sigma_x^2 = \frac{(\beta^2+1)^2}{\beta^4}\frac{L_x^2}{8\ln(2)N}\frac{(\beta^2+1)^2}{(\beta^2+2)^2}S_{x,PN} + \frac{(\beta^2+1)^2}{\beta^4}\frac{\sigma_{RON}^2}{d^2}\frac{\pi L_x^3 L_y}{128\ln^2(2)N^2}(\beta^2+1)^2 S_{x,RON}$$
$$\sigma_y^2 = \frac{(\beta^2+1)^2}{\beta^4}\frac{L_y^2}{8\ln(2)N}\frac{(\beta^2+1)^2}{(\beta^2+2)^2}S_{y,PN} + \frac{(\beta^2+1)^2}{\beta^4}\frac{\sigma_{RON}^2}{d^2}\frac{\pi L_y^3 L_x}{128\ln^2(2)N^2}(\beta^2+1)^2 S_{y,RON} \quad (7)$$

with the correction factors S involving error functions defined as:

$$S_{x,PN} = \frac{\mathbf{erf}\left(\frac{D\sqrt{\ln(2)(\beta^2+2)}}{\beta L_x}\right)\mathbf{erf}\left(\frac{D\sqrt{\ln(2)(\beta^2+2)}}{\beta L_y}\right)}{\mathbf{erf}^2\left(\frac{D\sqrt{\ln(2)(\beta^2+1)}}{\beta L_x}\right)\mathbf{erf}^2\left(\frac{D\sqrt{\ln(2)(\beta^2+1)}}{\beta L_y}\right)}\left[1 - \frac{2D\sqrt{\ln(2)(\beta^2+2)}e^{-\frac{\ln(2)(\beta^2+2)D^2}{\beta^2 L_x^2}}}{\sqrt{\pi}\beta L_x \mathbf{erf}\left(\frac{D\sqrt{\ln(2)(\beta^2+2)}}{\beta L_x}\right)}\right]$$
$$S_{y,PN} = \frac{\mathbf{erf}\left(\frac{D\sqrt{\ln(2)(\beta^2+2)}}{\beta L_x}\right)\mathbf{erf}\left(\frac{D\sqrt{\ln(2)(\beta^2+2)}}{\beta L_y}\right)}{\mathbf{erf}^2\left(\frac{D\sqrt{\ln(2)(\beta^2+1)}}{\beta L_x}\right)\mathbf{erf}^2\left(\frac{D\sqrt{\ln(2)(\beta^2+1)}}{\beta L_y}\right)}\left[1 - \frac{2D\sqrt{\ln(2)(\beta^2+2)}e^{-\frac{\ln(2)(\beta^2+2)D^2}{\beta^2 L_y^2}}}{\sqrt{\pi}\beta L_y \mathbf{erf}\left(\frac{D\sqrt{\ln(2)(\beta^2+2)}}{\beta L_y}\right)}\right]$$
$$S_{x,RON} = \frac{\mathbf{erf}\left(\frac{D\sqrt{2\ln(2)}}{\beta L_x}\right)\mathbf{erf}\left(\frac{D\sqrt{2\ln(2)}}{\beta L_y}\right)}{\mathbf{erf}^2\left(\frac{D\sqrt{\ln(2)(\beta^2+1)}}{\beta L_x}\right)\mathbf{erf}^2\left(\frac{D\sqrt{\ln(2)(\beta^2+1)}}{\beta L_y}\right)}\left[1 - \frac{2D\sqrt{2\ln(2)}e^{-\frac{2\ln(2)D^2}{\beta^2 L_x^2}}}{\sqrt{\pi}\beta L_x \mathbf{erf}\left(\frac{D\sqrt{2\ln(2)}}{\beta L_x}\right)}\right]$$
$$S_{y,RON} = \frac{\mathbf{erf}\left(\frac{D\sqrt{2\ln(2)}}{\beta L_x}\right)\mathbf{erf}\left(\frac{D\sqrt{2\ln(2)}}{\beta L_y}\right)}{\mathbf{erf}^2\left(\frac{D\sqrt{\ln(2)(\beta^2+1)}}{\beta L_x}\right)\mathbf{erf}^2\left(\frac{D\sqrt{\ln(2)(\beta^2+1)}}{\beta L_y}\right)}\left[1 - \frac{2D\sqrt{2\ln(2)}e^{-\frac{2\ln(2)D^2}{\beta^2 L_y^2}}}{\sqrt{\pi}\beta L_y \mathbf{erf}\left(\frac{D\sqrt{2\ln(2)}}{\beta L_y}\right)}\right] \quad (8)$$

where $\mathbf{erf}(x) = \frac{2}{\sqrt{\pi}}\int_0^x e^{-t^2}dt$ is the classical error function.

**2.3.2 Effect of flux**

Using simulated spot images with added photon and RON noise ($\sigma^2_{RON} = 2.5$ e⁻ per pixel[6]), CoG variance is measured over 10 000 Monte Carlo realizations for both ideal and real microlenses as a function of flux (non-elongated case, $L_x = L_\gamma = 1.5''$). Results are shown in Fig. 6 alongside the analytical curves from Eq. 6. There is good agreement with the theory, with a decrease in variance in $1/N$ at high flux and in $1/N^2$ at low flux (N being the number of photons per subaperture per frame). Differences come from the under sampling of the spots.

The real microlens increases CoG variance by a factor of 1.8 to 2.4 across the flux range. At nominal operating flux (N = 700– 1000 photons/subaperture/frame), approximately 1.9× more photons are required to achieve the same centroiding accuracy as with an ideal lens.

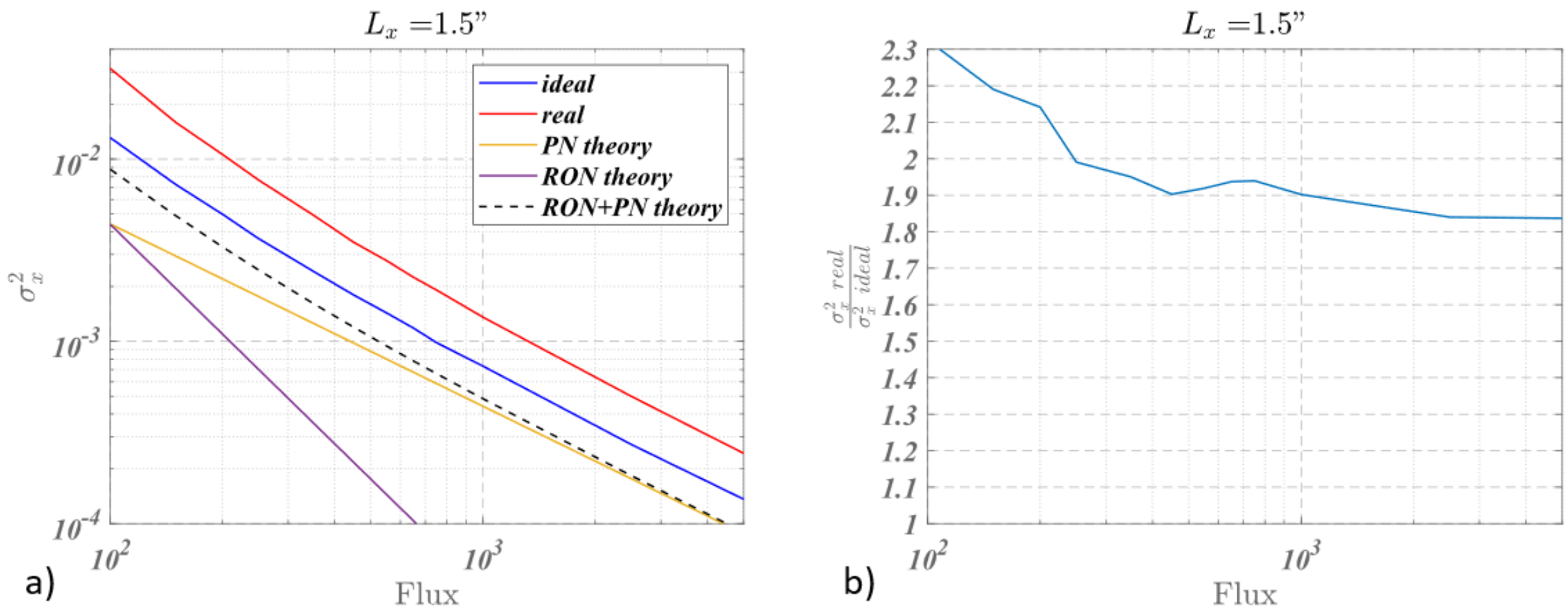

Fig 6 a) COG variance as a function of flux with an ideal lens and real lens. The LGS is non elongated (symmetric Gaussian map wth a FWHM of 1.5''). b) ratio between the variances of the real and the ideal lens

### 2.3.2 Effect of elongation

At fixed flux (N = 1000 photons/frame), CoG variance is evaluated as a function of LGS elongation along the x-axis ($L_\gamma$ fixed at 1.5″). To be consistent with the theoretical model, we start with a Gaussian spot as defined in equation (3). As shown in Fig.7, precision along the elongation axis decreases with increasing elongation, while the perpendicular axis also degrades due to the RON coupling term.

In Fig.8, the physically-modeled LGS sodium profile has been used and results are quite similar to those obtained with a Gaussian spot. Importantly, greater elongation reduces the relative impact of microlens shape errors: the ratio of real-to-ideal CoG variance decreases from ~2 at $L_x$ = 1.5″ to ~1.3–1.5 for $L_x$ > 8″ (Fig.8 c)). This occurs because the elongated spot is less sensitive to the central-peak flux redistribution induced by the real microlens profile.

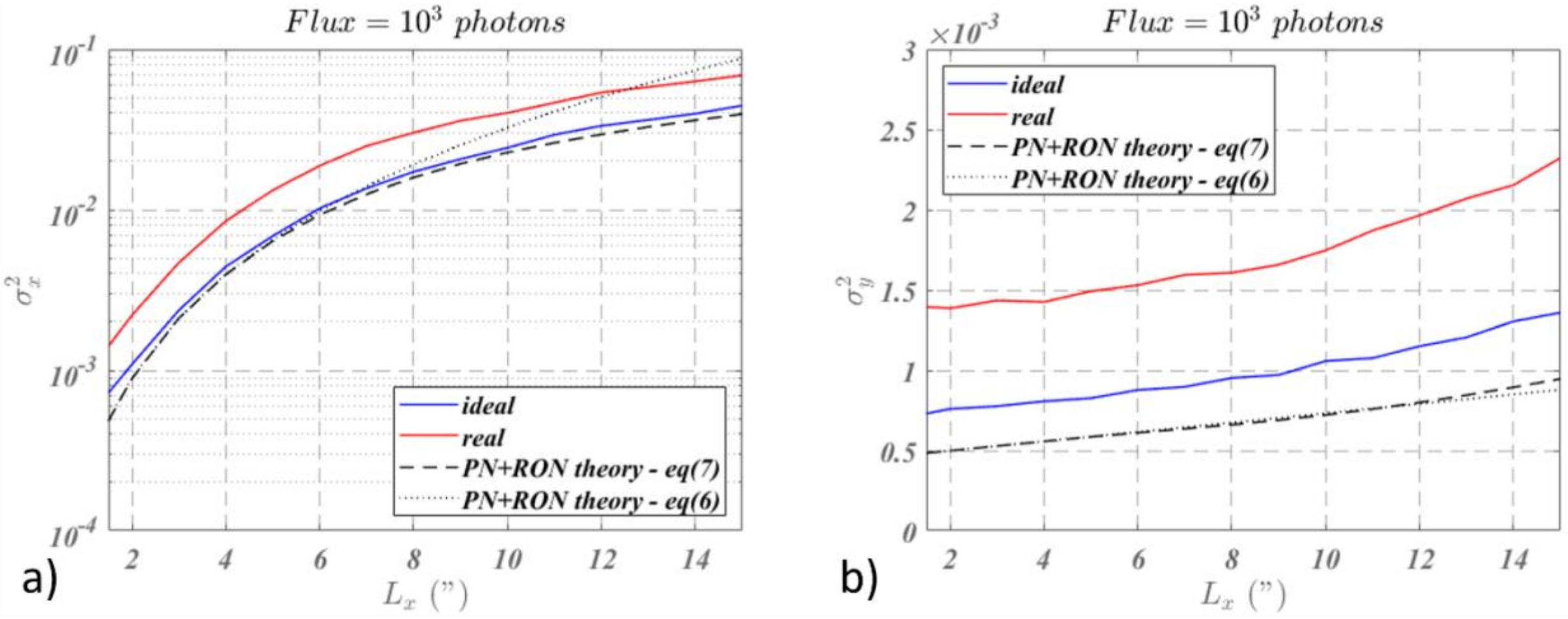

Fig 7. CoG variance as a function of x-axis elongation for a Gaussian spot, with ideal and real microlenses: (a) along the elongation axis; (b) along the perpendicular axis.

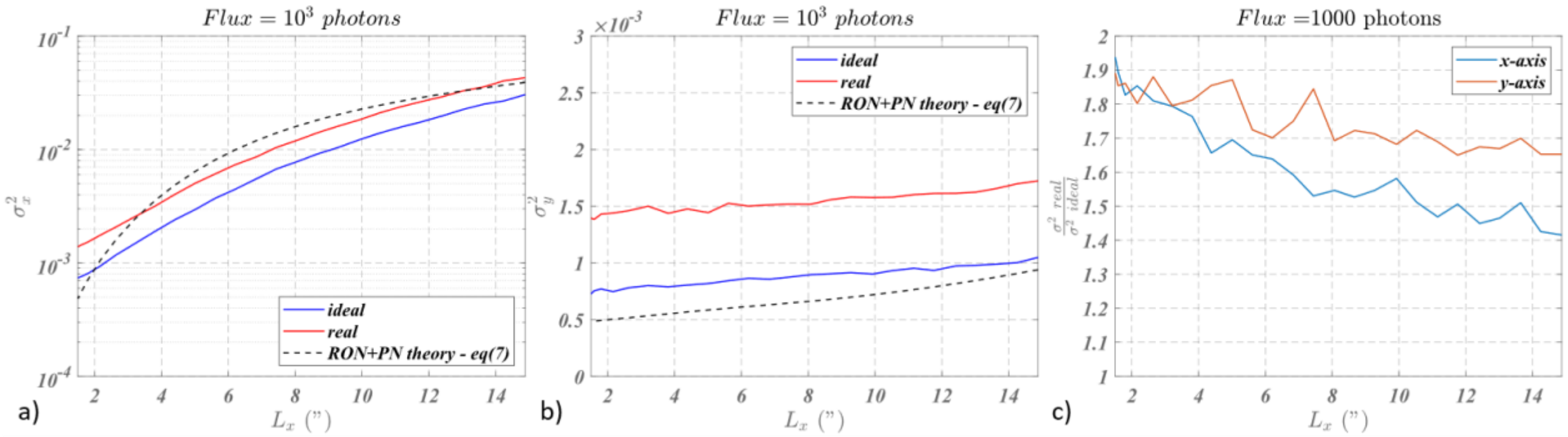

Fig 8. a), b) as Fig. 7 but using the physically-modeled LGS sodium profile. (c) Ratio of real-to-ideal CoG variances as a function of elongation.

### 2.3.3 Comparison of microlens prototypes

Three microlens prototypes (#1, #2, #3) drawn from different arrays are compared, with sag values of 5.71 μm, 6.53 μm, and 7.02 μm respectively. As sag increases, the real lens profile converges toward the ideal sphere: PSF wings decrease and the central peak flux increases (Fig. 9). Prototype #3 (largest sag) shows the smallest CoG variance penalty relative to the ideal lens. This provides a clear manufacturing specification: higher sag improves centroiding accuracy.

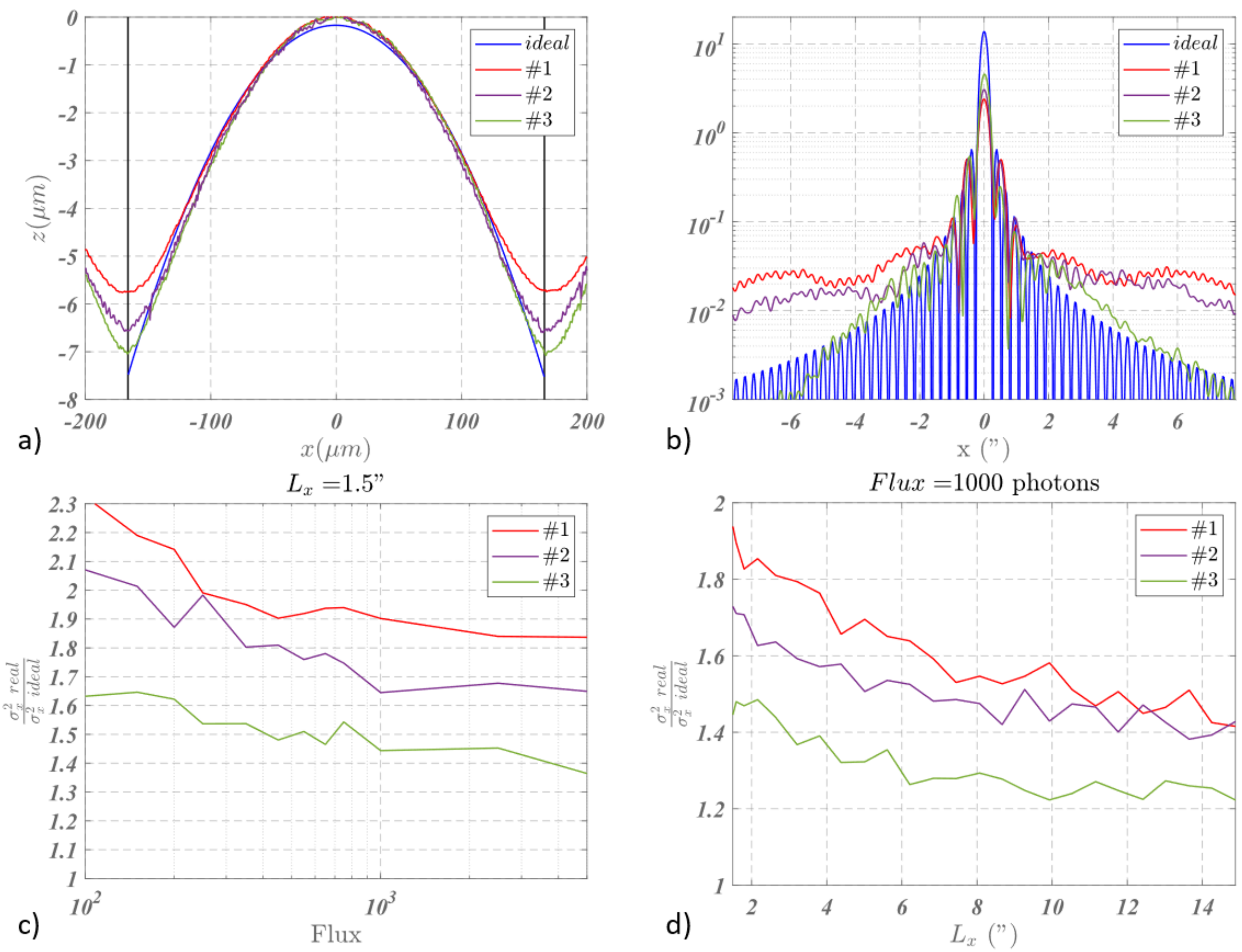

Fig 9. (a) Altitude profiles of three microlens prototypes along x-axis ($y = 0$). (b) Corresponding PSF profiles. (c) CoG variance ratio (real/ideal) as a function of flux at fixed elongation. (d) CoG variance ratio as a function of elongation at fixed flux.

## 3. IMPACT ON TOMOGRAPHIC ADAPTIVE OPTICS PERFORMANCE

We now quantify the impact of microlens shape imperfections on the overall AO performance of a six-LGS laser tomography AO (LTAO) system representative of ELT instruments. The simulation parameters, corresponding to the MORFEO LGS wavefront sensor configuration, are summarized in Table 1. The 39-m telescope pupil is sampled by 68×68 subapertures per LGS WFS, each 14×14 pixels of 1.15″ width. Atmosphere statistics correspond to the Cerro Armazones site profile ($r_0$ = 13.85 cm at 500 nm, $L_0$ = 25 m, zenith angle 30°). These parameters are representative of any ELT LTAO system, and the conclusions drawn here apply equally to future instruments on the ELT, GMT, or TMT.

Direct inclusion of measured microlens phase maps into the full simulation would require sampling all 4624 lenslets at high spatial resolution — computationally intractable for a 39-m telescope. Instead, we exploit the result of Section 2: in the photon-noise regime the CoG variance scales as $1/N$, so a real microlens is equivalent to an effective flux reduction at the subaperture level. The flux loss factor for each prototype is read from the high-flux asymptote of the CoG variance ratio curves (Fig. 9d) as a function of subaperture elongation. Each subaperture in the tomographic simulation is then assigned a throughput equal to the inverse of this factor.

The resulting K-band Strehl Ratio as a function of LGS flux is shown in Fig. 11 for a perfect microlens, prototype #3, and prototype #1. The effective flux penalty at the system level (1.25× for prototype #3, 1.4× for prototype #1) is significantly lower than single-subaperture predictions (1.5× and 2×, respectively). This mitigation arises from the redundancy of the tomographic measurement across six LGS and the ability of the MMSE reconstructor to downweight noisier subapertures via its noise covariance matrix.[7] While the overall impact remains within acceptable bounds, design margins are reduced, motivating continued improvement of microlens manufacturing.

Table 1. System configurations for tomographic AO end-to-end simulations (representative of the MORFEO LGS wavefront sensor configuration on the ELT).

| System | Parameter | Value |
|---|---|---|
| Telescope | Diameter | 39 m |
| | AO Pupil | ELT pupil (no spiders) |
| | Imager Pupil | Circular |
| Atmosphere | Layers | 9 layers from 35-layer Cerro Armazones profile |
| | $r_0$ @ 500 nm | 13.85 cm |
| | $L_0$ | 25 m |
| | Zenith Angle | 30 deg |
| LGS | Number | 6 |
| | Asterism Radius | 34 arcsec |
| | Na Profile | Uniform (18 km thick at 90 km) |
| | Wavelength | 589 nm |
| HO-WFS | Number | 6 |
| | Subapertures | 68×68 |
| | Pixel scale | 14×14 pixels × 1.15″/pixel |
| | Type | Diffractive WFS |
| | Flux regime | Variable |

| LO-WFS | Number | 1 |
|---|---|---|
| | Subapertures | 2×2 for Tip/Tilt |
| | Type | Geometric |
| | Flux Regime | Noiseless |

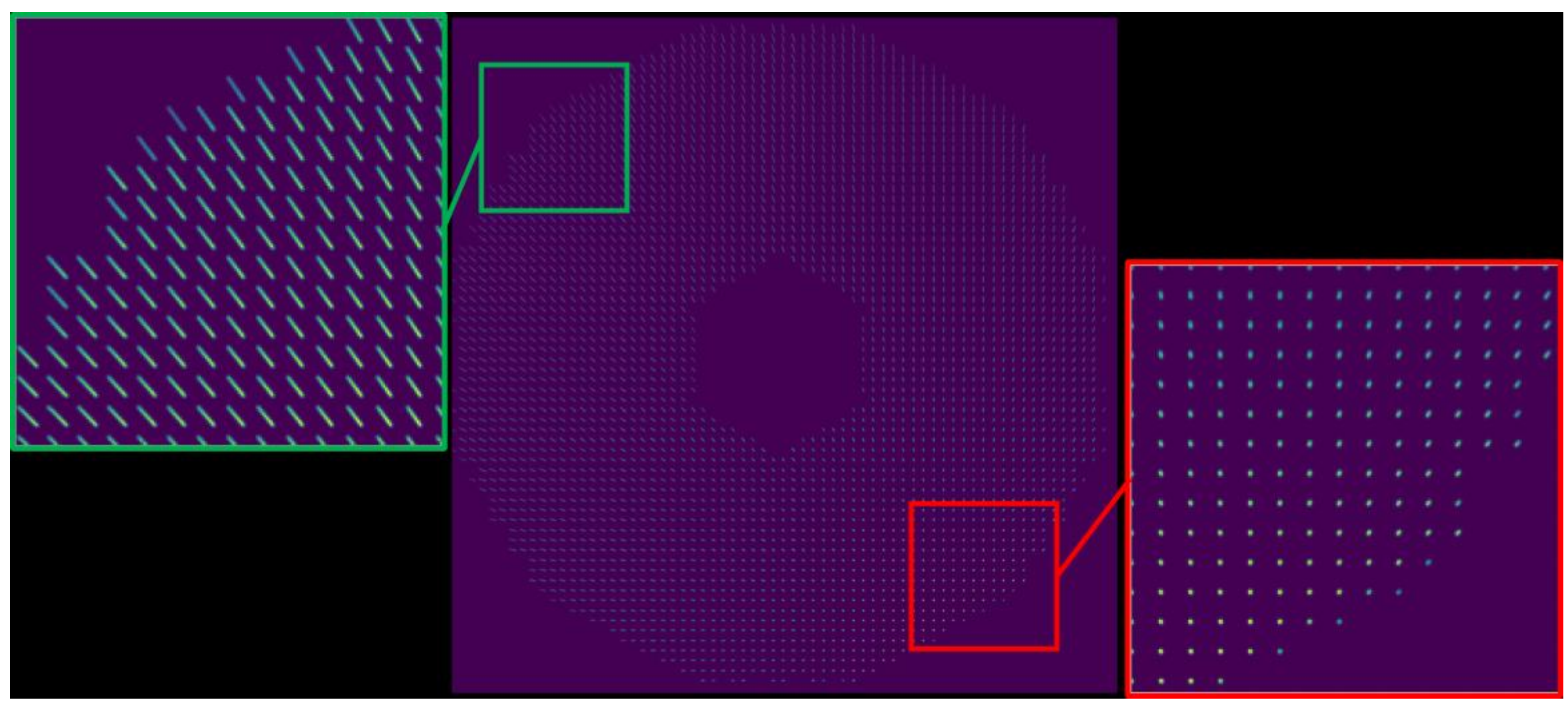

Fig 10. LGS spot elongation pattern across the 68×68 SH WFS pupil. Elongation ranges from near-zero at the pupil center to ~14″ at the edges.

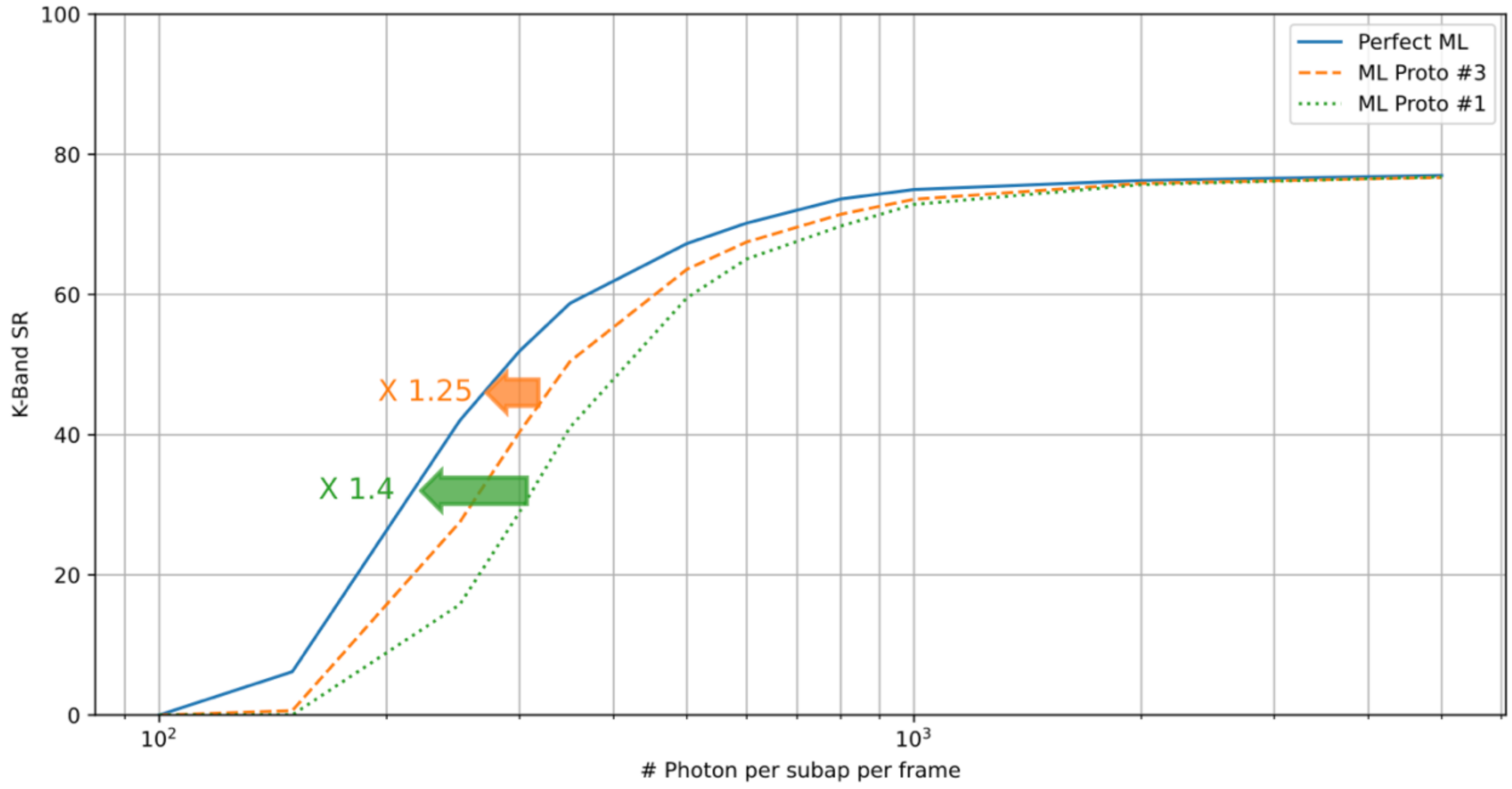

Fig. 11 — K-band Strehl Ratio as a function of LGS flux per subaperture per frame for a perfect microlens (solid blue), prototype #3 (dashed orange), and prototype #1 (dotted green). Horizontal arrows indicate the effective flux loss in terms of system-level AO performance.

# 4. EXPERIMENTAL VALIDATION WITH THE LGS WFS BENCH AT LAM

An optical bench prototype of the MORFEO/HARMONI LGS wavefront sensor has been constructed at LAM to test and validate the WFS design.[12] Beyond validating the simulations of Section 2, this bench enables characterization of the entire microlens array in a single measurement — an important advantage given that profiling all 4624 lenslets individually with the Wyko interferometer would be virtually impossible. The methodology described here is generic and applicable to any SH-WFS array characterization campaign.

## 4.1 Bench description

The bench comprises three main subsystems: a source module, a Spatial Light Modulator (SLM), and a laser detection module (LDM).[1] The source is a 400 µm-core multi-mode LED fiber source, equivalent to a 1″ LGS on sky. Light from the source is collimated by a lens (f = 45 mm, D = 24 mm) acting as the entrance pupil. The SLM, conjugated to the pupil, can emulate a deformable mirror or introduce phase aberrations; for this study it is used as a flat mirror. Light is then relayed to the microlens array via an afocal system, and the C-BLUE camera (characterized in Ref. 6) is placed in the focal plane.

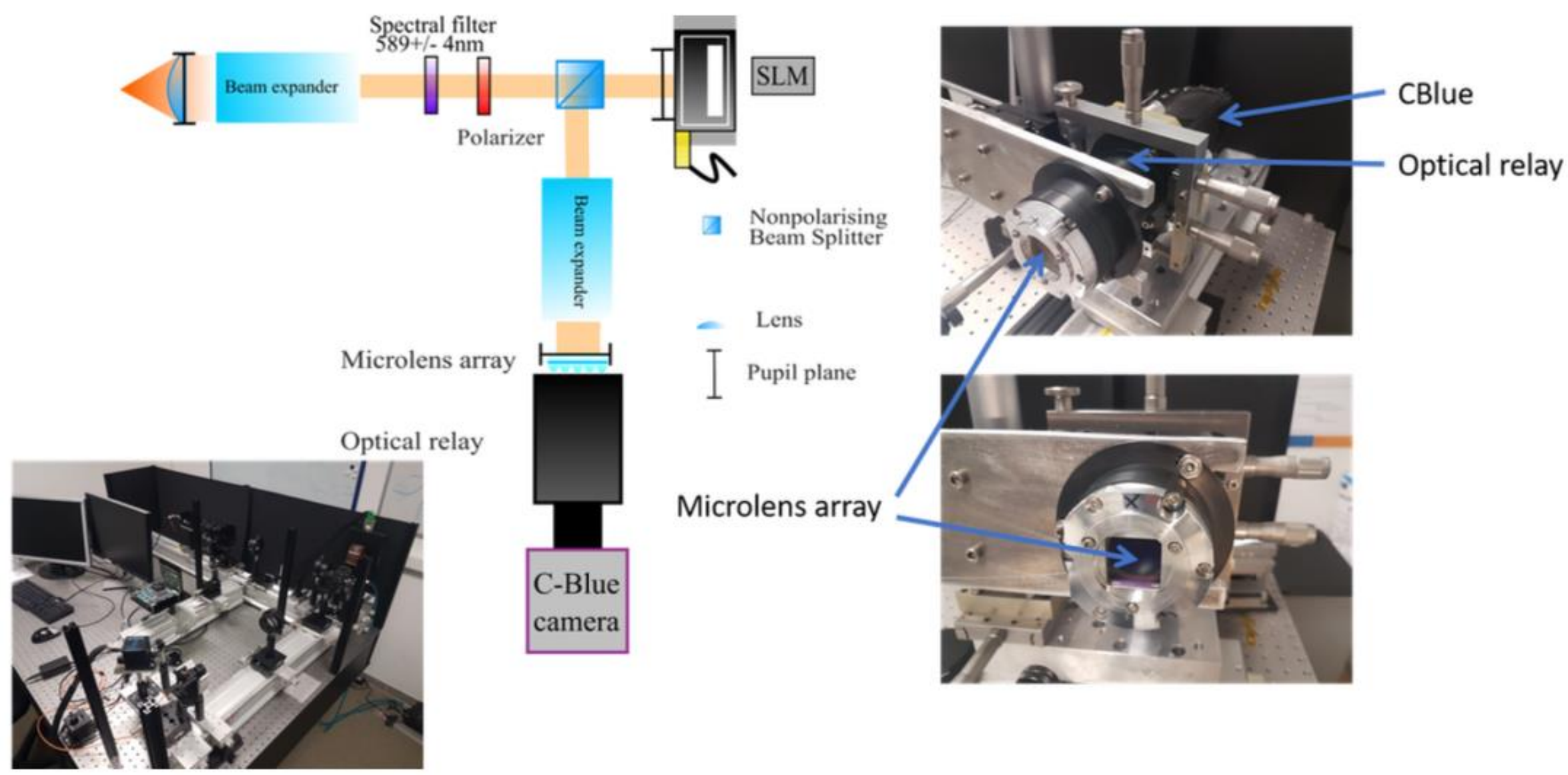

Fig 12. LGS wavefront sensor optical bench at LAM. Main subsystems: beam expander, spectral filter (589 nm), polarizer, SLM, afocal relay, microlens array, optical relay, and C-BLUE detector.

## 4.2 Full-array characterization method

Camera frames are recorded at 500 Hz, consistent with the planned operating frame rate. The photon count per pixel is derived as $N^{ph} = G \times ADU / QE$, with $G = 0.24\ e^-/ADU$ and $QE = 0.7\ e^-/ph$.[6] For each subaperture, the FWHM is measured at high flux and the WCoG variance computed over 1000 frames as a function of flux.

To isolate the microlens contribution, we compare measured CoG variance against the theoretical variance expected for a Gaussian spot of the same measured flux and FWHM. The ratio $R = \sigma^{2Measured} / \sigma^{2Gauss}$ quantifies the excess variance attributable to microlens shape errors. Figures 13–15 show camera images, single-subaperture validation, and the spatial maps of flux, FWHM, and CoG variance across the pupil. Figure 16 shows the resulting ratio maps and histograms for three prototype arrays.

Prototype #2 (purple histogram in Fig. 16b, median ratio = 1.81) yields the best performance overall, with the lowest median ratio and smallest inter-lenslet scatter. These results are consistent with the sag-dependent trend identified in Section 2.3.3 and confirm the validity of the simulation-based characterization framework.

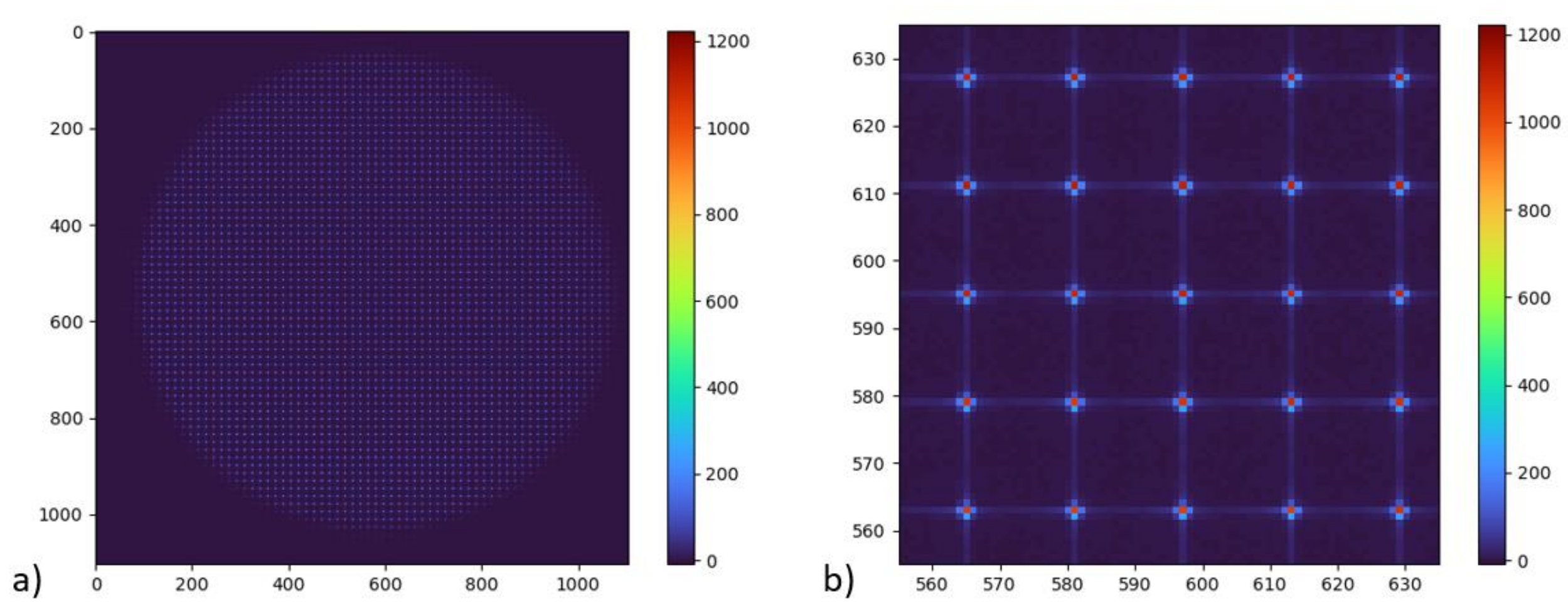

Fig 13. C-BLUE camera images: (a) full detector showing the complete pupil (68×68 subapertures); (b) zoom on a 5×5 sub-array.

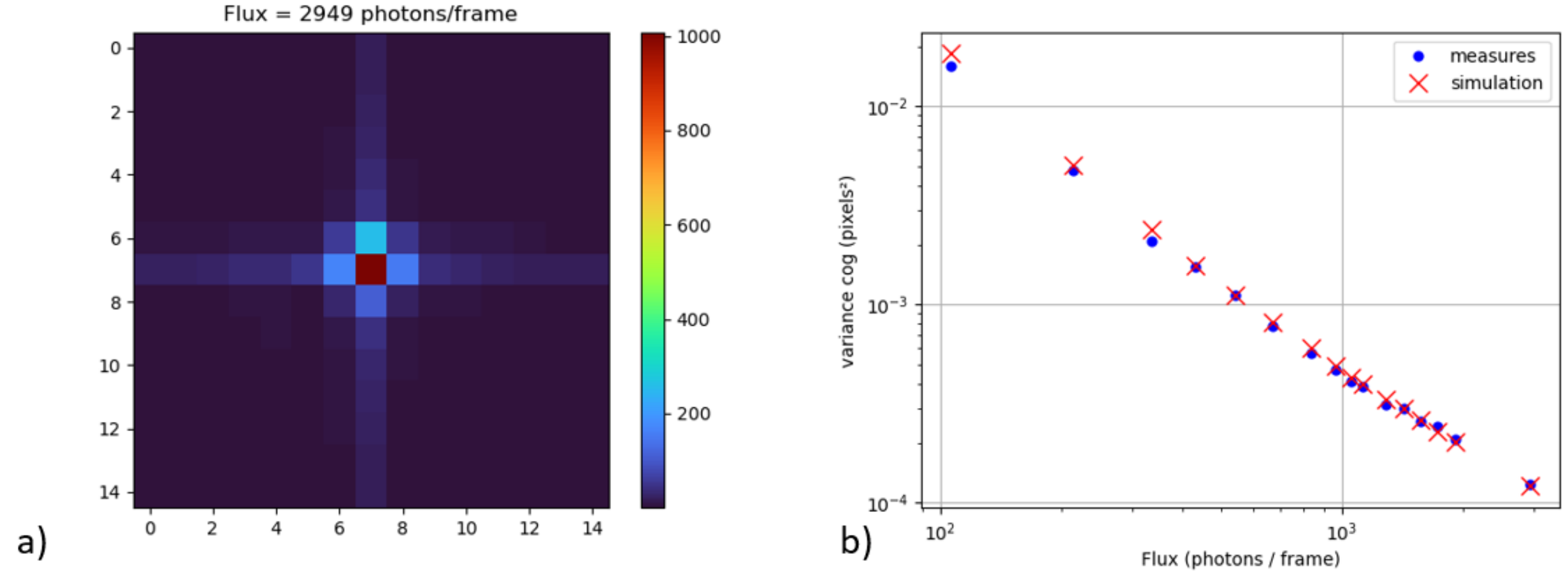

Fig 14. (a) Mean image of one subaperture averaged over 1000 frames at high flux. (b) CoG variance vs. flux for that subaperture: measurements (blue dots) vs. simulation (red crosses).

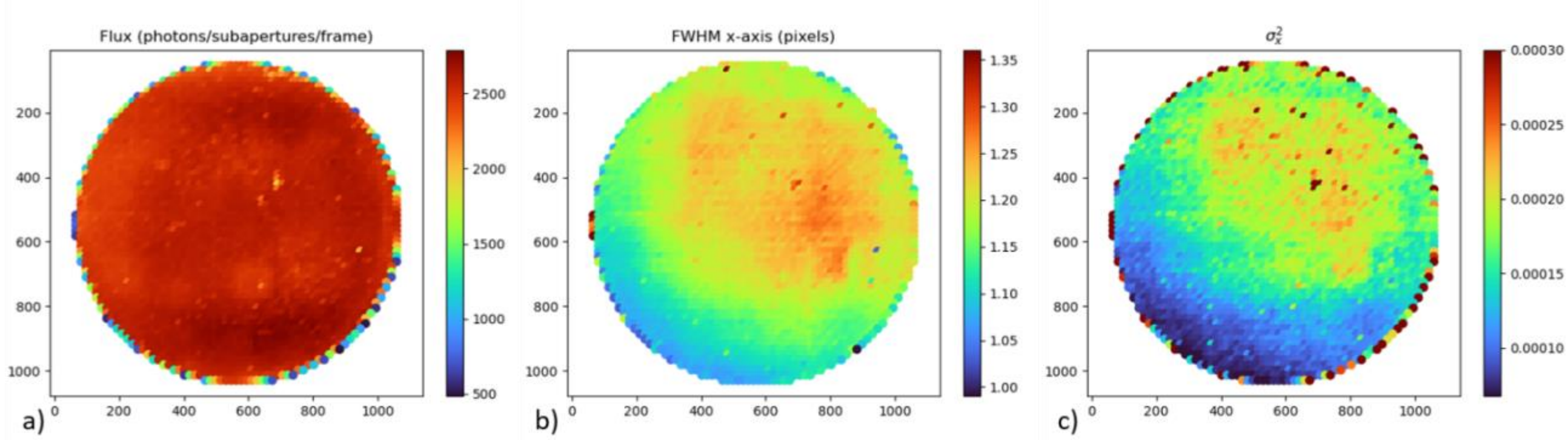

Fig 15. Pupil maps (one dot per subaperture): (a) flux (photons/subaperture/frame); (b) FWHM along x-axis; (c) measured CoG variance.

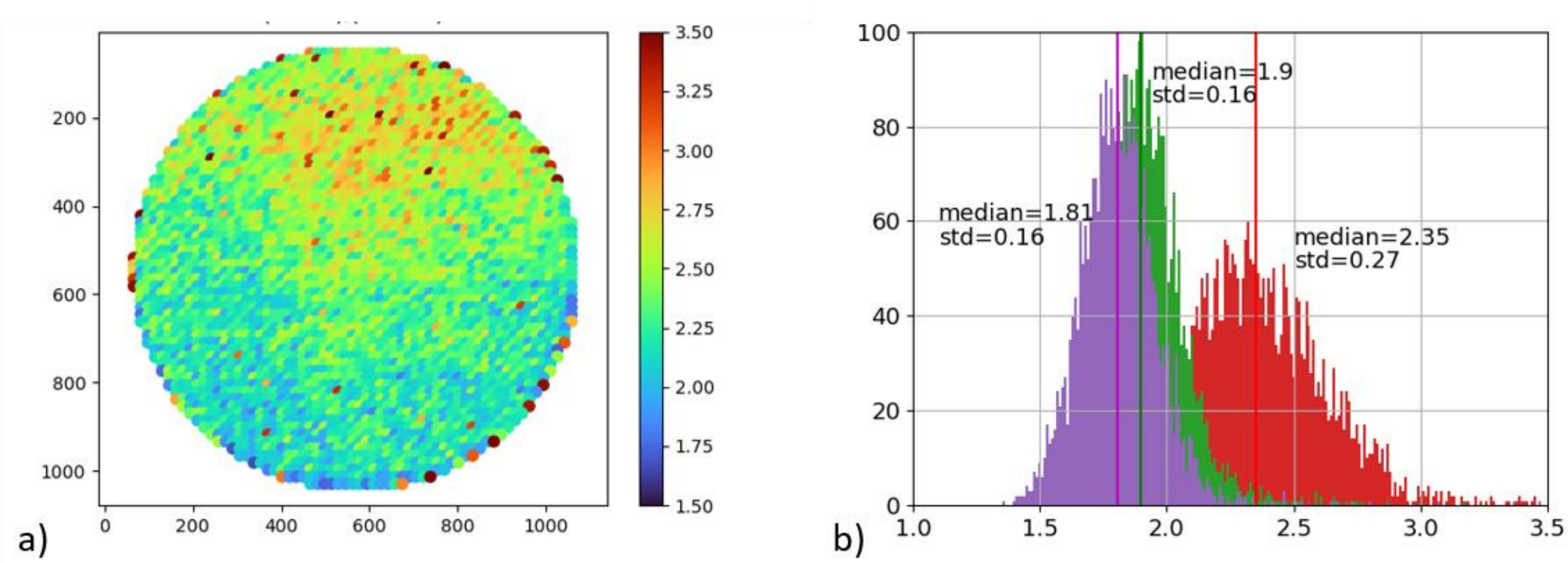

Fig 16. Excess CoG variance ratio (measured/theoretical) for three microlens arrays: (a) spatial map across the pupil; (b) histograms for arrays #1 (red, median = 2.35), #2 (purple, median = 1.81), and #3 (green, median = 1.9).

## 5. CONCLUSION

We have presented a comprehensive characterization of the impact of microlens surface shape imperfections on the performance of LGS Shack-Hartmann wavefront sensors for ELT-class telescopes. The study was conducted in the context of the MORFEO LGS wavefront sensor, but the methodology and results are broadly applicable to any similar system, including future LGS instruments on the ELT, GMT, and TMT.

An optical model was developed from interferometric surface measurements, allowing the CoG variance penalty to be quantified as a function of flux, LGS elongation, and lens shape. The main findings are: (i) a real microlens degrades CoG variance by 1.8–2.4× relative to an ideal lens, equivalent to roughly doubling the required photon flux; (ii) this penalty decreases with increasing LGS elongation, as the elongated spot is less sensitive to central-peak flux redistribution; and (iii) higher microlens sag improves centroiding accuracy and reduces the CoG variance penalty.

A computationally efficient method was developed to propagate per-subaperture results into full tomographic AO simulations by modeling microlens errors as effective subaperture flux losses. System-level simulations demonstrate that the MMSE tomographic reconstructor, exploiting measurement redundancy and elongation diversity across six LGS, significantly mitigates the performance penalty: effective K-band SR flux losses of 1.25–1.4× at the system level, compared to 1.5–2.0× at the individual subaperture level.

Experimental validation using the LGS WFS bench at LAM confirmed the simulation results and demonstrated a new full-array characterization method based on per-subaperture CoG variance ratios. New microlens arrays are currently being manufactured; the tools developed here will be used to qualify them. The methodology is broadly applicable to any Shack-Hartmann system operating in a low-flux regime.

## ACKNOWLEDGMENTS

This work benefited from the support of the French National Research Agency (ANR) through grants WOLF (ANR-18-CE31-0018), APPLY (ANR-19-CE31-0011), and LabEx FOCUS (ANR-11-LABX-0013); the Programme Investissement Avenir F-CELT (ANR-21-ESRE-0008); the Action Spécifique Haute Résolution Angulaire (ASHRA) of CNRS/INSU co-funded by CNES; the Région Sud; and the French government under the France 2030 investment plan as part of the Initiative d'Excellence d'Aix-Marseille Université – A*MIDEX (AMX-22-RE-AB-151).